# Detection of lightning in Saturn's Northern Hemisphere

Mohsen H. Moghimi

Department of Aerospace Engineering, Sharif University of Technology, Tehran, Iran;
m_moghimi@alum.sharif.edu

**Abstract** During Cassini flyby of Saturn at a radial distance $6.18 R_s$ (Saturn Radius), a signal was detected from about 200 to 430 Hz that had the proper dispersion characteristics to be a whistler. The frequency-time dispersion of the whistler was found to be 81 Hz1/2s. Based on this dispersion constant, we determined, from a travel time computation, that the whistler must have originated from lightning in the northern hemisphere of Saturn. Using a simple centrifugal potential model consisting of water group ions, and hydrogen ions we also determine the fractional concentration and scale height that gave the best fit to the observed dispersion.

## 1 Introduction

Whistlers are electromagnetic waves which have frequencies below the electron cyclotron frequency and electron plasma frequency. Lightning also produce this wave and make it possible to propagate along the magnetic field lines from the source of lightning to the detector. In 1935 Eckersley carried out a detailed study of whistler and show that the time delay varied as $\frac{D}{\sqrt{f}}$, where $D$ is dispersion constant and f is frequency. This relation is called the Eckersley law.

He could not explain the long delay time, which was sometimes several seconds or more. In 1953 Storey was the first to confirm that whistlers were produced by lightning. He observed a lightning flash, and heard the whistling tone a few seconds later. Storey was also the first to propose the correct theoretical explanation of whistlers. He showed that whistlers propagate along the Earth's magnetic field lines from one hemisphere to the other in a plasma mode of propagation. We also assumed parallel propagation in the analysis of the whistler. The signal produced by a lightning flash is converted into a whistling tone as it propagates along the magnetic field line, with the highest frequencies arriving first.

During the previous space missions to the outer planet of the solar system, whistlers have been detected at two other planets. In 1979 during the Voyager I flyby of Jupiter Whistler wave has been detected. Later during the Voyager I and Voyager II flybys of Saturn this wave could not be found. It is possible that due to the limitation of the obtained data during the mentioned flybys Whistler could not be detected. Also no Whistler detected during the Voyager II flyby of Uranus in 1986. But the successful detection of this wave has been occurred during the Voyager II flyby of Neptune in 1989. In this article detection of Whistler from lightning during the Cassini flyby of Saturn in October 28th, 2004 has been investigated.

## 2 Observation

A frequency-time spectrogram of an event which is believed to be a Whistler is shown in Fig.1. This event is observed between 09:58:52 UT and 09:58:55 UT. At this time spacecraft was at the northern hemisphere with latitude of 12.31 degrees and radial distance of $6.18 R_s$. The duration of the whistler was about 3 seconds and the frequency varied from 430 to 200 Hz. Because the group velocity of whistler mode waves is greater at higher frequencies than at lower frequencies, the higher frequencies are detected first. This property is called dispersion.

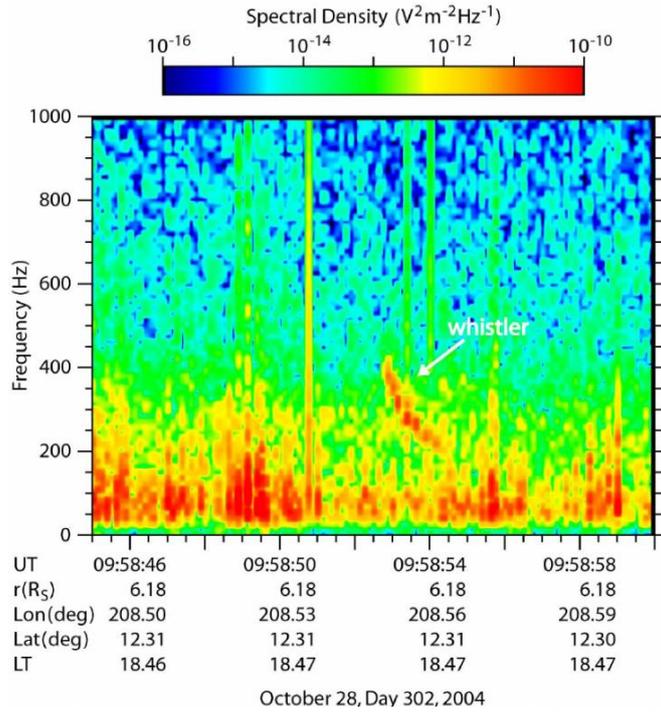

**Fig1.** A frequency-time spectrogram showing the strongest and clearest whistler detected.

## 3   Investigation of the location of the lightning

To locate the lightning a very useful dispersion relation which Eckersley introduced (Ref.1) in 1935 was applied. This relation has following form:

$$t = t_0 + \frac{D}{\sqrt{f}} \quad (1)$$

where $t$ is the arrival time of the Whistler at frequency $f$, $t_0$ is the time of the lightning, and $D$ is dispersion constant. Fig.2 is a plot of frequency versus time which has been derived from the given data in the spectrogram of Fig.1.

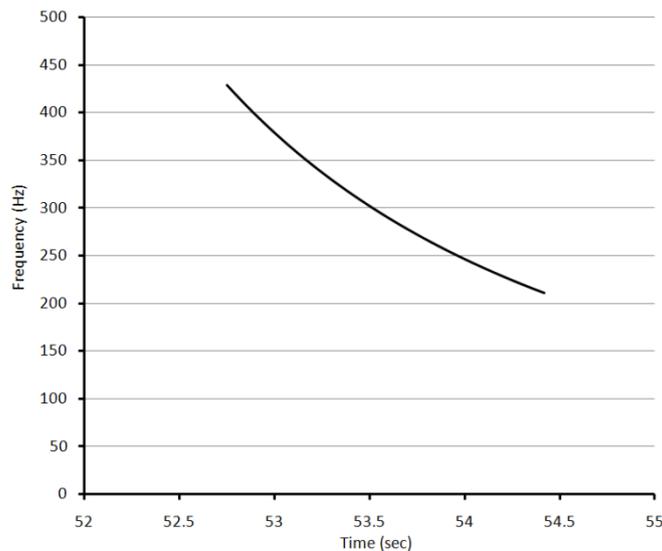

**Fig2.** Frequency values of the whistler are plotted as a function of time.

To calculate the dispersion constant we should plot $\frac{1}{\sqrt{f}}$ versus the time. If the Eckersley law is obeyed then the plot must be a straight line. Fig.3 shows the obtained plot which is linear. The slop of the graph is equal to the inverse of the dispersion constant. Regarding to this issue the dispersion constant in Eq. (1), $D$, is about 81.

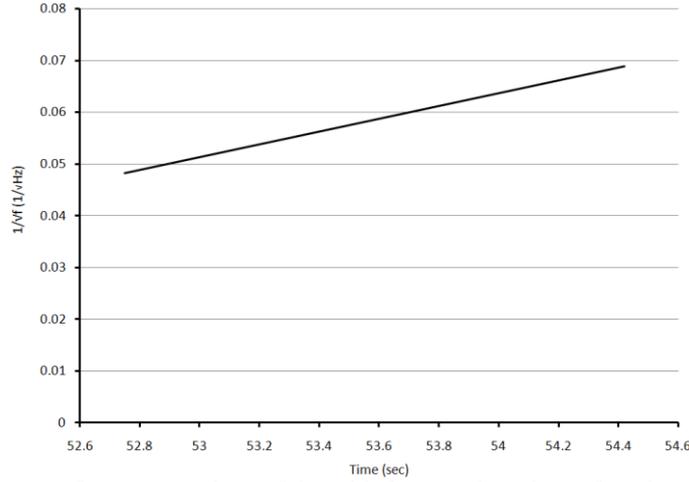
**Fig3.** Inverse frequency values of the whistler are plotted as a function of time.

As mentioned in the introduction section Storey assumed that the Whistler wave will propagate along the magnetic field line. The plasma in Saturn's magnetosphere beyond the rings is known to be rotating as though it is rigidly locked to Saturn's rotation. This rotation produces a centrifugal force that causes the plasma to accumulate in a disk near the equatorial plane. This disk is called the plasma disk.

To achieve the location of the lightning we should calculate the analytical dispersion constant and compare it to the obtained dispersion constant from the data. In addition, since the propagation velocity of the Whistler mode depends on the electron density it is evident that measurements of the Whistler dispersion can provide important information on the plasma density distribution in the outer magnetosphere. Then this goal could be achieved by an analytical relation which has been introduced by Helliwell in 1965 (Ref.2). For parallel propagation of the wave along the field lines the general equation for the arrival time of whistler is as the following:

$$t = \frac{1}{c} \int \frac{1 + \left(\frac{1}{2}\right)\left[\frac{f_p^2 f_c}{f(f_c - f)^2}\right]}{\left[1 + \frac{f_p^2}{f(f_c - f)}\right]^{1/2}} ds \quad (2)$$

where $c$ is the speed of light, $f_c$ is the electron cyclotron frequency, $f_p$ is the electron plasma frequency, and $s$ is the magnetic field line path. Because the ionosphere is very thin only the negligible dispersion occurs as the Whistler passes through the ionosphere. Above the ionosphere, in the region where the cyclotron frequency is very large and the plasma density is very small ($f \ll f_c$ and $f f_c \geq f_p^2$) the group index of refraction $n_g = \frac{1 + \left(\frac{1}{2}\right)\left[\frac{f_p^2 f_c}{f(f_c - f)^2}\right]}{\left[1 + \frac{f_p^2}{f(f_c - f)}\right]^{1/2}}$ is close to one, so very little dispersion occurs in this regions. As the Whistler propagates farther away from the planet, the cyclotron frequency decreases until the condition $f f_c \geq f_p^2$ is no longer satisfied. At this point the group index of refraction starts to increase and approaches the following form in the limit $f f_c \ll f_p^2$ (The assumption of $f \ll f_c$ is as the previous.):

$$n_g \cong \frac{1}{2} \frac{f_p}{\sqrt{f f_c}} \quad (3)$$

This condition, $f f_c \ll f_p^2$, dictates us that the dispersion is important in this region. Also the $\frac{1}{\sqrt{f}}$ dependence of travel time makes a similarity between the Eckersley law and the formula for arrival time. Then by substituting Eq. (3) in Eq. (2) the dispersion relation is as the following:

$$D = \frac{1}{2c} \int \frac{f_p}{\sqrt{f_c}} ds \quad (4)$$

The relation to calculate $f_c$ is $28B$ ($B$ is the magnetic field strength in nT); then a magnetic field model is required to evaluate this parameter. Since the deviation of the Saturn's magnetic dipole is about one degree from its rotational axis then it is possible to use the magnetic field model for a dipole which has been introduced by Parks in 1991 (Ref.3). The relation for this model is as the following:

$$B = B_0 \frac{\sqrt{1 + 3sin^2\lambda}}{cos^6\lambda} \tag{5}$$

where $B_0$ is the magnetic field strength at the equator, and $\lambda$ is the latitude. To evaluate $f_p$ the relation $8980\sqrt{n_e}$ should be applied. In this relation $n_e$ is the electron particle density and the unite is $cm^{-3}$. Both of the units for these frequencies are in Hz.

It is clear from the relation of electron plasma frequency that electron particle density should be calculated. To obtain a simple model for electron density in the Saturn's magnetosphere we assume that the dominant force beyond $2.3R_s$ is the centrifugal force acting on co-rotating plasma. The centrifugal force in cylindrical coordinate could be written as below:

$$F = m_i r\omega^2 \tag{6}$$

where $m_i$ is the ion mass, $r$ is the perpendicular distance from the rotational axis of the planet, and $\omega$ is the rotation rate of the Saturn. The potential energy of a particle in the co-rotating plasma can be derived by integrating the centrifugal force as the following:

$$W = -\int F dr = -\int m_i r\omega^2 dr = -\frac{1}{2}m_i r^2 \omega^2 \tag{7}$$

Because the spacecraft Cassini and the Saturn are in a spherical system then a relation in the form of $r = Rcos\lambda$ should be replaced in the Eq. (6). In this relation $R$ is the redial distance from the center of the Saturn. The relation between this parameter and the point $R_0$ at which the magnetic field line intersects the equatorial plane is $R = R_0 cos^2\lambda$. Then by substituting these relations in Eq. (6), the potential energy of the ion moving along the magnetic field line of a dipole can be expressed as below:

$$W = -\frac{1}{2}m_i R_0^2 cos^6\lambda \omega^2 \tag{8}$$

To specify the equatorial plane as a potential energy reference it is beneficial to subtract above equation from the potential energy at this plane. Then the potential energy has the following form:

$$W = \frac{1}{2}m_i \omega^2 R_0^2 (1 - cos^6\lambda) \tag{9}$$

From statistical mechanics the number density of ion can be expressed as below:

$$n_i = \sum_i \alpha_i n_0 exp\left[-\frac{m_i \omega^2 R_0^2}{2KT_i}(1 - cos^6\lambda)\right] \tag{10}$$

where $\alpha_i = \frac{n_{i0}}{n_0}$ is the fractional concentration of the ion species, $K$ is Boltzmann's constant, $n_0$ is the ion density at $\lambda = 0$, and $T_i$ is the ion temperature. From the quasi-neutrality condition of the plasma the electron number density is equal to ion number density; then Eq. (10) can be written for electron too. In addition, the term $\frac{2KT_i}{3m_i\omega^2}$ is plasma scale height ($H_i$).

$$n_e = \sum_i \alpha_i n_0 exp\left[-\frac{1}{3}\frac{R_0^2}{H_i^2}(1 - cos^6\lambda)\right] \tag{11}$$

If the plasma consists of a single species of ions and an equal density of electrons then Eq. (11) can be shown as the following:

$$n_e = n_0 exp\left[-\frac{1}{3}\frac{R_0^2}{H^2}(1 - cos^6\lambda)\right] \tag{12}$$

At this point we proceed to evaluate the Eq. (4) for the northern and southern hemisphere. This integral can be evaluated using below expression for the path line (Ref.3).

$$\frac{ds}{d\lambda} = R_0 cos\lambda\sqrt{1 + 3sin^2\lambda} \tag{13}$$

This equation gives us a relation between the arc length of a dipole field lines and the latitude. To calculate $R_0$ and the latitude of the foot of the magnetic field line, $\lambda_0$, we proceed as the following:

$$R = R_0 cos^2\lambda \Rightarrow 6.18R_s = R_0 cos^2 12.31 \Rightarrow R_0 = 6.47R_s \tag{14}$$

$$\lambda_0 = cos^{-1}\sqrt{\frac{1}{R_0}} = cos^{-1}\sqrt{\frac{1}{6.47}} \Rightarrow \lambda_0 = 66.85°$$

The measured magnetic field at the spacecraft latitude is 89.513 nT; then by substituting the spacecraft latitude which is 12.31 degrees, in Eq. (5) the magnetic field at the equator, $B_0$, is 73.0 nT. The calculation is as below:

$$B = B_0 \frac{\sqrt{1+3sin^2\lambda}}{cos^6\lambda} \Rightarrow 89.513 = B_0 \frac{\sqrt{1+3sin^2 12.31}}{cos^6 12.31} \Rightarrow B_0 = 73.0\ nT \quad (15)$$

The equatorial electron density in Eq. (12) was determined by requiring that the electron density at the spacecraft latitude is $7.5\ cm^{-3}$ which determined from measurements. For scale height ranging from 1 to 4 the dispersion calculated for both northern and southern hemisphere. For northern hemisphere (Fig.4) the dispersion varies from 30 to 130 $\sqrt{Hz}\ sec$ and for the southern hemisphere varies from about 250 to 315 $\sqrt{Hz}\ sec$. Then the dispersion with value 81 $\sqrt{Hz}\ sec$ is in northern hemisphere and the related scale height is about $1.92R_s$.

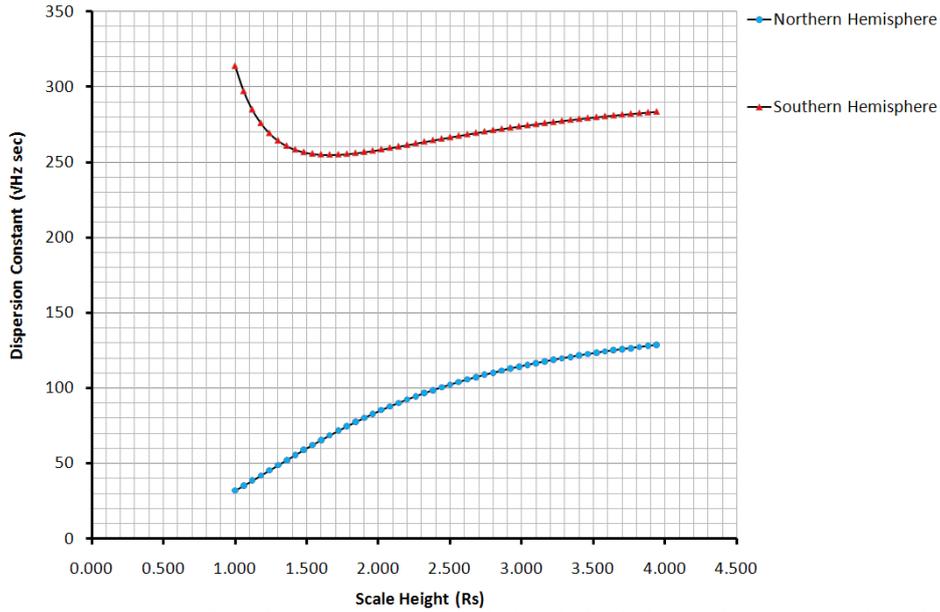

**Fig4.** Dispersion constant is plotted as a function of scale height both for northern and southern hemisphere.

## 4 Electron Density Distribution Considering Water Group Ions and Protons

Plasma measurements on the Cassini spacecraft show that the plasma disk consists of hydrogen ions (protons) and water group ions ($OH^+, H_2O^+, H_3O^+$) (Ref.4). In this section we use Eq. (11) and Eq. (2). The reason of Eq. (2) instead of Eq. (4) is: Eq. (4) fails for high latitudes, where the electron number density is low; then the violating condition $ff_c \ll f_p^2$. Also it is clear by reviewing Eq. (11). If the latitude increases continuously the electron number density will decrease inversely. The fractional concentrations in Eq. (11) are $\alpha_{H^+}$ and $\alpha_{W^+}$ respectively. Also the scale heights for these two species are $H_{H^+}$ and $H_{W^+}$ respectively. The extended form of Eq. (11) is as the following:

$$n_e = \alpha_{H^+} n_0 exp\left[-\frac{1}{3}\frac{R_0^2}{H_{H^+}^2}(1-cos^6\lambda)\right] + \alpha_{W^+} n_0 exp\left[-\frac{1}{3}\frac{R_0^2}{H_{W^+}^2}(1-cos^6\lambda)\right] \quad (16)$$

The unknowns of this equation are $\alpha_{H^+}$, $\alpha_{W^+}$, $H_{H^+}$, and $H_{W^+}$ respectively. The relation between fractional concentrations is $\alpha_{H^+} + \alpha_{W^+} = 1$ or the equivalent relation is $n_{H^+} + n_{W^+} = n_0$. The value of $n_0$ derived by RPWS measurements of the upper hybrid resonance frequency ($f_{UH}$) during five passes through the inner magnetosphere from October 2004 to March 2005. Fig.5 is the spectrogram for March 8[th], 2005 to March 10[th], 2005.

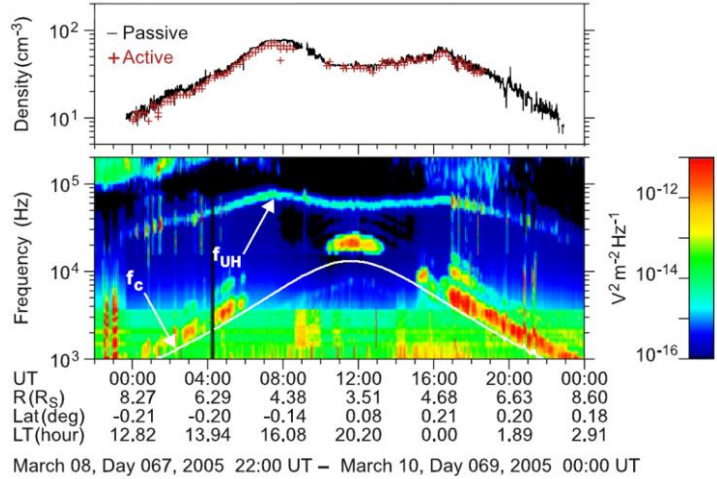

**Fig5.** The frequency-time spectrogram in the lower panel shows the electric field intensities detected by the Cassini RPWS during the pass through Saturn's inner magnetosphere on March 8–10, 2005. The upper panel shows the electron densities. (Ref.5)

By using relation $n_e = \frac{f_{UH}^2 - f_c^2}{(8980)^2}$ it is possible to calculate $n_0$. The calculation shows this value is 25 $cm^{-3}$. To this point there are three constraints to four unknowns in Eq. (16). The constraints are dispersion constant, electron density at spacecraft latitude, and electron density at equatorial plane for the $R_0 = 6.47 R_s$. The fourth constraint could be achieved by defining parameter $\beta$ that is the ratio of the scale height for the protons to the scale height if the water group ions ($\beta = \frac{H_{H^+}}{H_{W^+}}$). This ration can be expressed as $\beta = 4\sqrt{\frac{T_{H^+}}{T_{W^+}}}$. The coefficient 4 is due to the mass ratio of the water group ions to the protons which is 16. The best measurement of the ratio $\frac{T_{H^+}}{T_{W^+}}$ is from Richardson which has been delivered in 1998. Richardson (Ref.6) showed that the ration $\frac{T_{H^+}}{T_{W^+}}$ is between values 0.16 to 0.5. Then $\beta$ varies between 1.6 to 2.8. By estimating the ratio and then calculating the fractional concentrations, the electron density will be obtained. Then the electron plasma frequency can be derived and the arrival time can be calculated. By using an error method (for example mean squared error) it is possible to decrease the errors in each estimation step. The calculations show that the best answer can be achieved by $\beta$ varies about 3 to 4. The differences between some calculated times and the actual times has been showed in Fig.6.

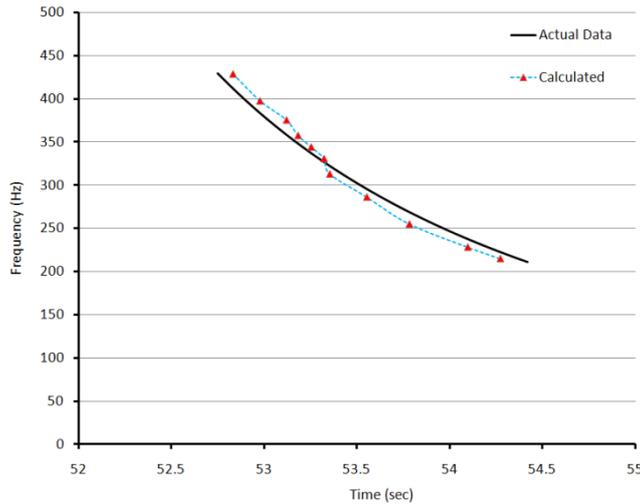

**Fig6.** Frequency versus time plot for actual time and the calculated time.

The full electron density equation and the four calculated parameters for $\beta = 4$ are: $\alpha_{H^+} = 0.17, H_{H^+} = 4.12 R_s, H_{W^+} = 1.03 R_s$. The plot of the total electron particle density and electron density for each of species versus the path length has been shown in Fig.7. The violet lines show the position of the spacecraft and its electron density number. The vertical violet line intersects the two lines for species electron number distribution lines approximately at the same electron number density. This shows that species contributions have the same value at spacecraft position.

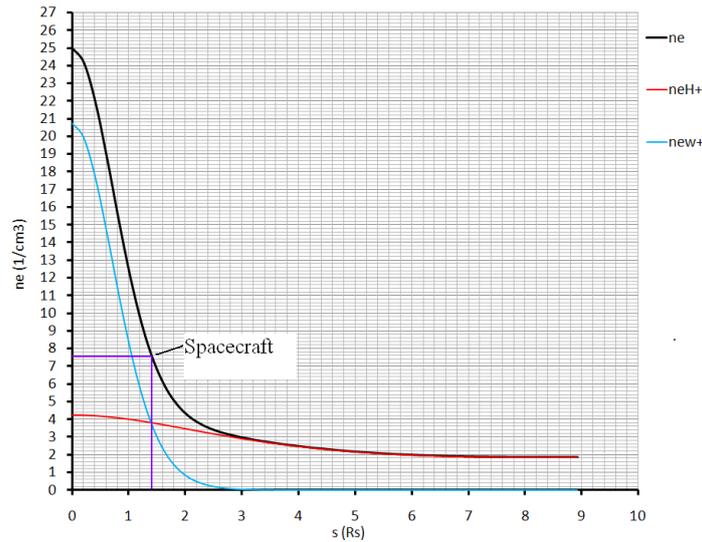

**Fig7.** Electron Number density versus arc length to show the concentration fraction of hydrogen group and water group ions of plasma along the field line.

## 5  Conclusion

To detect Whistlers the spacecraft must be relatively close to the Saturn, otherwise the electron cyclotron frequency would be too small for the Whistler to propagate to the spacecraft. Since the value for dispersion constant in northern hemisphere is equal to the obtained dispersion constant from the Cassini spectrogram, it is believed that the lightning have occurred in this hemisphere and at latitude 66.85 degrees. Saturn's Electrical Discharges which have been detected in September 2004 indicated lightings occurred in Saturn due to a storm around 35S latitude. Time sequence of this storm was showed in Fig.8.

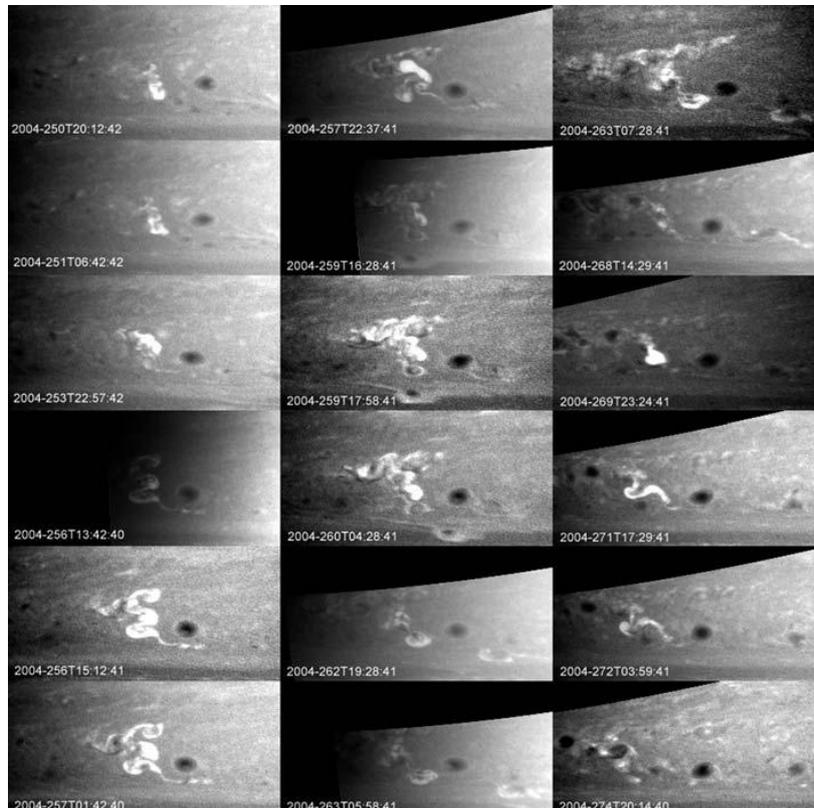

**Fig8.** Time sequence of the September storm at 35-S latitude. The sequence begins on day 250 (6 September 2004) and ends on day 274. The storm reaches its peak intensity in visible light on days 255 to 260. After fading for a few days, it returns as a high-contrast bright spot on day 269. Between day 272 and day 274, a dark filament appears and is observed curling up on its way to becoming a new dark spot (Ref.7).

This latitude is in the southern hemisphere and near the equatorial region; but the detected lightning in October 2004 has latitude around 66.85 degrees in northern hemisphere. Then even if this storm lasted until the detection date of the whistler it is unlikely to be a reason for this detection because of the big difference in latitudes. In addition, due to the tilt of Saturn's rotational axis this high northern latitude will be almost in darkness. This issue described that sunlight cannot be a source for convective activity at that latitude and occurrence of the lightning might be due to an internal energy source.

The model used for electron density consists of two groups of ions. The protons and the water group ions. These species defining is regarding to the research which has been presented by Young in 2005. As it is clear from the electron number density this model has a good accuracy near the equatorial plane where some effects like gravitational effects, ambipolar effects and anisotropy are relatively small. But for future researches at high latitude this items must be considered to obtain an accurate model for the electron number density. However, dispersion constant showed that it is possible to calculate the particle distribution with constraints due to the nature of the whistler's waves.